\documentclass[english,aps,superscriptaddress]{revtex4}
\usepackage[T1]{fontenc}
\usepackage[latin9]{inputenc}
\usepackage{verbatim}
\usepackage{amsbsy}
\usepackage{amssymb}
\usepackage{graphicx}
\usepackage{esint}

\makeatletter
\@ifundefined{textcolor}{}
{%
 \definecolor{BLACK}{gray}{0}
 \definecolor{WHITE}{gray}{1}
 \definecolor{RED}{rgb}{1,0,0}
 \definecolor{GREEN}{rgb}{0,1,0}
 \definecolor{BLUE}{rgb}{0,0,1}
 \definecolor{CYAN}{cmyk}{1,0,0,0}
 \definecolor{MAGENTA}{cmyk}{0,1,0,0}
 \definecolor{YELLOW}{cmyk}{0,0,1,0}
}

\makeatother

\usepackage{babel}
\begin{document}
\title{What can we learn from global spin alignment of $\phi$ meson in heavy-ion
collisions?}
\author{Xin-Li Sheng}
\affiliation{Department of Modern Physics, University of Science and Technology
of China, Hefei, Anhui 230026, China}
\affiliation{Institute for Theoretical Physics, Goethe University of Frankfurt,
Max-von-Laue-Str. 1, D-60438 Frankfurt am Main, Germany}
\author{Lucia Oliva}
\affiliation{Institute for Theoretical Physics, Goethe University of Frankfurt,
Max-von-Laue-Str. 1, D-60438 Frankfurt am Main, Germany}
\author{Qun Wang}
\affiliation{Department of Modern Physics, University of Science and Technology
of China, Hefei, Anhui 230026, China}
\begin{abstract}
We propose that a significant positive deviation from 1/3 for the
spin density matrix element $\rho_{00}$ of the $\phi$ meson may
indicate the existence of a mean field of the $\phi$ meson generated
in heavy-ion collisions. This explains why STAR preliminary data for
the $\phi$ meson's $\rho_{00}$ are much larger than $1/3$ while
the data of $\Lambda$ and $\bar{\Lambda}$ polarization seem not
to allow such a significant and positive deviation. The contribution
may be from the polarization of the strange quark and antiquark through
the $\phi$ field, an effective mode of the gluon field in strong
interaction. We show that $\rho_{00}$ for the $\phi$ meson is a
good analyzer for fields even if they may strongly fluctuate in space-time.
\end{abstract}
\maketitle

\section{Introduction}

The rotation and spin polarization are inherently correlated and can
be converted from one to another in materials as manifested in the
Barnett effect \cite{Barnett:1935} and the Einstein-de Haas effect
\cite{dehaas:1915}. One of the most recent examples is that an electric
voltage from the spin-current is observed to be generated from the
vortical motion in a liquid metal \cite{Takahashi:2016}. In ultra-relativistic
heavy-ion collisions (HIC), a huge orbital angular momentum (OAM)
can also be generated mainly along the direction perpendicular to
the reaction plane \cite{Liang:2004ph,Liang:2004xn,Voloshin:2004ha,Betz:2007kg,Becattini:2007sr,Gao:2007bc}
(see, e.g. \cite{Wang:2017jpl}, for a recent review). Such a huge
OAM is distributed into the hot and dense quark matter and converted
to global polarization of hadrons through the spin-orbit coupling
\cite{Liang:2004ph,Gao:2007bc,Zhang:2019xya} in a microscopic approach
or spin-vorticity coupling in a macroscopic approach \cite{Becattini:2013fla,Becattini:2016gvu,Fang:2016vpj,Pang:2016igs,Florkowski:2017dyn,Florkowski:2018ahw}.
The STAR collaboration has recently measured a non-vanishing global
polarization of $\Lambda$ hyperons in Au+Au collisions at $\sqrt{s_{NN}}=7.7-200$
GeV \cite{STAR:2017ckg,Adam:2018ivw}.

Accompanying a huge OAM in HIC, a strong magnetic field is also formed,
pointing to the same direction \cite{Skokov:2009qp,Voronyuk:2011jd,Toneev:2011aa,Deng:2012pc,McLerran:2013hla,Gursoy:2014aka,Tuchin:2014iua,Li:2016tel}.
The OAM and magnetic field lead to chiral effects of massless fermions:
the chiral magnetic effect (CME) which probes the topological fluctuation
of quantum chromodynamics vacuum \cite{Vilenkin:1980fu,Kharzeev:2007jp,Fukushima:2008xe}
(see, e.g. \cite{Kharzeev:2015znc}, for a recent review) and the
chiral vortical effect (CVE) \cite{Vilenkin:1978hb,Erdmenger:2008rm,Banerjee:2008th,Son:2009tf,Gao:2012ix,Hou:2012xg,Gao:2015zka,Gao:2018jsi}
which probes the vorticity field of the fluid. One of the most active
research in HIC experiments is to search for the CME \cite{Abelev:2009ac,Abelev:2009ad,Abelev:2012pa,Adamczyk:2013hsi,Adamczyk:2013kcb,Adamczyk:2014mzf,Khachatryan:2016got,Sirunyan:2017quh,Acharya:2017fau}.
However, the CME has not been observed due to dominant backgrounds.
Furthermore, no direct and definite effects from electric and magnetic
fields have been found so far. The challenge comes from the fact that
the lifetime of the electric and magnetic field is so short ($\lesssim$1
fm/c) that they can be regarded as a pulse.

While the polarization of $\Lambda$ can be measured by its weak decay,
the polarization of vector mesons cannot be measured since they mainly
decay through strong interaction. However, the spin alignment of a
vector meson can only be measured through $\rho_{00}$, the 00-element
of its spin density matrix, encoded in the angular distribution of
its decay daughters \cite{Liang:2004xn,Yang:2017sdk}. If $\rho_{00}\neq1/3$,
the distribution is anisotropic and the spin of the vector meson is
aligned to the spin quantization direction. In 2008, the STAR collaboration
measured $\rho_{00}$ for the vector meson $\phi(1020)$ in Au+Au
collisions at 200 GeV, which is consistent to $1/3$ indicating no
spin alignment within errors \cite{Abelev:2008ag}. Recent STAR's
preliminary data for the $\phi$ meson's $\rho_{00}$ or $\rho_{00}^{\phi}$
at lower energies show a significant POSITIVE deviation from $1/3$,
which is far beyond our current understanding of the polarization
\cite{Zhou:2017nwi}. In this note, we will show that such a large
POSITIVE deviation of $\rho_{00}^{\phi}$ from 1/3 may imply the existence
of a mean field for the $\phi$ meson in heavy ion collisions.  

\section{Conventional understanding for spin alignment of $\phi$ meson}

The 00-element of the spin density matrix $\rho_{00}$ for the vector
meson enters the angular distribution of its decay daughter as 
\begin{equation}
\frac{dN}{d\cos\theta}=\frac{3}{4}\left[(1-\rho_{00})+(3\rho_{00}-1)\cos^{2}\theta\right],
\end{equation}
where $\theta$ is the angle between the daughter's momentum and the
spin quantization direction \cite{Liang:2004xn,Yang:2017sdk}. The
STAR preliminary data imply that $\rho_{00}^{\phi}>1/3$ and significantly
deviate from 1/3. In the coalescence or combination model the $s$
and $\bar{s}$ quark form a $\phi$ meson, and $\rho_{00}^{\phi}$
is related to the polarization $P_{s}$ and $P_{\bar{s}}$ for $s$
and $\bar{s}$ respectively, 
\begin{equation}
\rho_{00}^{\phi}\approx\frac{1}{3}-\frac{4}{9}P_{s}P_{\bar{s}},\label{eq:rho-00-1}
\end{equation}
if $P_{s}$ and $P_{\bar{s}}$ are both small. In a simple model,
the spin polarization of $\Lambda$ and $\bar{\Lambda}$ is carried
by $s$ and $\bar{s}$ respectively, so we have $P_{\Lambda}=P_{s}$
and $P_{\bar{\Lambda}}=P_{\bar{s}}$. Therefore $\rho_{00}^{\phi}$
in (\ref{eq:rho-00-1}) is approximately 
\begin{equation}
\rho_{00}^{\phi}\approx\frac{1}{3}-\frac{4}{9}P_{\Lambda}P_{\bar{\Lambda}}\lesssim\frac{1}{3},\label{eq:rho-00-phi-13}
\end{equation}
where $P_{\Lambda}$ and $P_{\bar{\Lambda}}$ can be estimated by
using the STAR data $P_{\Lambda}\approx(1.08\pm0.15\pm0.11)\%$ and
$P_{\bar{\Lambda}}\approx(1.38\pm0.30\pm0.13)\%$ \cite{STAR:2017ckg,Adam:2018ivw}:
$(4/9)P_{\Lambda}P_{\bar{\Lambda}}\approx6.6\times10^{-5}$. So the
STAR data for $P_{\Lambda}$ and $P_{\bar{\Lambda}}$ seem to imply
that $\rho_{00}^{\phi}$ cannot be significantly larger than 1/3,
which contradicts the STAR preliminary data on $\rho_{00}^{\phi}$.
We will show that the key to reconcile such a conflict is that $P_{s}$
and $P_{\bar{s}}$ will have additional contributions which have never
been considered before. 

\section{Spin polarization in vorticity and electromagnetic field}

\label{sec:em}We take $xz$ plane as the reaction plane with one
nucleus moving along $+z$ direction at $x=-b/2$ while the other
nucleus moving along $-z$ direction at $x=b/2$. The OAM is along
$+y$ direction. 

From Eq. (64) in Ref. \cite{Yang:2017sdk} the spin polarization vector
(normalized to 1) for massive fermions (upper sign) and anti-fermions
(lower sign) in the vorticity and electromagnetic field is 
\begin{eqnarray}
P_{\pm}^{\mu}(x,p) & = & \frac{1}{2m}\left(\tilde{\omega}_{\mathrm{th}}^{\mu\nu}\pm\frac{1}{E_{p}T}Q\tilde{F}^{\mu\nu}\right)p_{\nu}\left[1-f_{FD}(E_{p}\mp\mu)\right],\label{eq:spin-polar-yang}
\end{eqnarray}
where $Q$ is the electric charge of the fermion, $p^{\mu}=(E_{p},\mathbf{p})$
denotes the four-momentum for fermion or anti-fermion with $E_{p}\equiv\sqrt{\mathbf{p}^{2}+m^{2}}$
being the energy of the fermion or anti-fermion, $\tilde{\omega}_{\mathrm{th}}^{\mu\nu}=\frac{1}{2}\epsilon^{\mu\nu\sigma\rho}\omega_{\sigma\rho}^{\mathrm{th}}$
is the dual thermal vorticity tensor with the thermal vorticity tensor
given by $\omega_{\sigma\rho}^{\mathrm{th}}=\frac{1}{2}[\partial_{\sigma}(\beta u_{\rho})-\partial_{\rho}(\beta u_{\sigma})]$
with $\beta\equiv1/T$, $\tilde{F}^{\mu\nu}=\frac{1}{2}\epsilon^{\mu\nu\sigma\rho}F_{\sigma\rho}$
is the dual electromagnetic field strength tensor, and $f_{FD}$ is
the Fermi-Dirac distribution. The electric and magnetic field as three-vectors
are defined as $E^{i}=E_{i}=F^{i0}$ and $B^{i}=B_{i}=-\frac{1}{2}\epsilon_{ijk}F^{jk}$
with $i,j,k=x,y,z$. In a similar way, one can define the thermal
vorticity three-vector $\omega^{i}=\omega_{i}=\tilde{\omega}_{\mathrm{th}}^{i0}$,
the 'magnetic' part of the thermal vorticity tensor, and the 'electric'
part of the thermal vorticity tensor $\varepsilon^{i}=\varepsilon_{i}=\omega_{\mathrm{th}}^{i0}$,
which are $\boldsymbol{\omega}=\frac{1}{2}\nabla\times(\beta\mathbf{u})$
and $\boldsymbol{\varepsilon}=-(1/2)[\partial_{t}(\beta\mathbf{u})+\nabla(\beta u^{0})]$
in three-vector forms. 

Applying Eq. (\ref{eq:spin-polar-yang}) to the strange and anti-strange
quark $s$ and $\bar{s}$, we obtain the polarization along the $y$
direction 
\begin{eqnarray}
P_{s/\bar{s}}^{y}(t,\mathbf{x},\mathbf{p}_{s/\bar{s}}) & = & \frac{1}{2}\omega_{y}+\frac{1}{2m_{s}}\hat{\mathbf{y}}\cdot(\boldsymbol{\varepsilon}\times\mathbf{p}_{s/\bar{s}})\nonumber \\
 &  & \pm\frac{Q_{s}}{2m_{s}T}B_{y}\pm\frac{Q_{s}}{2m_{s}^{2}T}\hat{\mathbf{y}}\cdot\left(\mathbf{E}\times\mathbf{p}_{s/\bar{s}}\right),\label{eq:py-s-sbar}
\end{eqnarray}
where $Q_{s}=-e/3$ is the electric charge of the $s$ quark ($e>0$),
and we have taken the non-relativistic limit $E_{p}\simeq m_{s}$
and the Boltzmann limit $1-f_{FD}(E_{p}\mp\mu)\simeq1$. The last
term of Eq. (\ref{eq:py-s-sbar}) is the spin-orbit term for quarks
in electric fields, the similar term is the key to the nuclear shell
structure if applying to nucleons in meson fields \cite{Mayer:1949pd,Haxel:1949fjd}. 

In the coalescence model, the polarization of $\Lambda$ or $\bar{\Lambda}$
in its rest frame is given by \cite{Yang:2017sdk} 
\begin{eqnarray}
P_{\Lambda/\bar{\Lambda}}^{y}(t,\mathbf{x}) & = & \frac{1}{3}\int\frac{d^{3}\mathbf{r}}{(2\pi)^{3}}\frac{d^{3}\mathbf{q}}{(2\pi)^{3}}\left|\psi_{\Lambda/\bar{\Lambda}}(\mathbf{q},\mathbf{r})\right|^{2}\nonumber \\
 &  & \times\left[P_{s/\bar{s}}^{y}(t,\mathbf{x},\mathbf{p}_{1})+P_{s/\bar{s}}^{y}(t,\mathbf{x},\mathbf{p}_{2})+P_{s/\bar{s}}^{y}(t,\mathbf{x},\mathbf{p}_{3})\right]\nonumber \\
 & = & \frac{1}{2}\omega_{y}\pm\frac{Q_{s}}{2m_{s}T}B_{y},\label{eq:p-lambda}
\end{eqnarray}
where $\psi_{\Lambda/\bar{\Lambda}}(\mathbf{q},\mathbf{r})$ are wave-functions
of $\Lambda/\bar{\Lambda}$ in momentum space with the normalization
condition $\int d^{3}\mathbf{r}d^{3}\mathbf{q}\left|\psi_{\Lambda/\bar{\Lambda}}(\mathbf{q},\mathbf{r})\right|^{2}=(2\pi)^{6}$,
and internal momenta of three quarks are denoted as $\mathbf{p}_{1}=\mathbf{r}/2+\mathbf{q}$,
$\mathbf{p}_{2}=\mathbf{r}/2-\mathbf{q}$ and $\mathbf{p}_{3}=-\mathbf{r}$
which satisfy $\mathbf{p}_{1}+\mathbf{p}_{2}+\mathbf{p}_{3}=0$ in
the rest frame of $\Lambda/\bar{\Lambda}$. In the square bracket
of Eq. (\ref{eq:p-lambda}), $P_{s/\bar{s}}^{y}(t,\mathbf{x},\mathbf{p}_{1})$
means that $\mathbf{p}_{1}$ is the momentum of the $s/\overline{s}$
quark in $\Lambda/\bar{\Lambda}$ (the momenta of two light quarks/antiquarks
are then $\mathbf{p}_{2}$ and $\mathbf{p}_{3}$), and $P_{s/\bar{s}}^{y}(t,\mathbf{x},\mathbf{p}_{2})$
and $P_{s/\bar{s}}^{y}(t,\mathbf{x},\mathbf{p}_{3})$ have similar
meanings. Comparing Eq. (\ref{eq:p-lambda}) with Eq. (\ref{eq:py-s-sbar}),
we see that there are no contributions from $\boldsymbol{\varepsilon}$
and $\mathbf{E}$ in $P_{\Lambda/\bar{\Lambda}}^{y}$. The reason
is that both $\boldsymbol{\varepsilon}$ and $\mathbf{E}$ terms in
$P_{s/\bar{s}}^{y}$ are linearly proportional to $\mathbf{p}$, so
these terms in the square bracket of Eq. (\ref{eq:p-lambda}) are
vanishing due to $\mathbf{p}_{1}+\mathbf{p}_{2}+\mathbf{p}_{3}=0$
in the rest frame of $\Lambda$ and $\bar{\Lambda}$. 

The 00-element of the spin density matrix for the $\phi$ meson is
calculated by \cite{Yang:2017sdk} 
\begin{eqnarray}
\rho_{00}^{\phi}(t,\mathbf{x}) & \approx & \frac{1}{3}-\frac{4}{9}\int\frac{d^{3}\mathbf{p}}{(2\pi)^{3}}P_{s}^{y}(\mathbf{p})P_{\bar{s}}^{y}(-\mathbf{p})\left|\psi_{\phi}(\mathbf{p})\right|^{2},\label{eq:rho00}
\end{eqnarray}
where $\psi_{\phi}(\mathbf{p})$ is the wavefunction in momentum space
for the $\phi$ meson with the normalization $\int d^{3}\mathbf{p}\left|\psi_{\phi}(\mathbf{p})\right|^{2}=(2\pi)^{3}$,
and we have put $\mathbf{p}_{s}=\mathbf{p}$ and $\mathbf{p}_{\bar{s}}=-\mathbf{p}$
in the center of mass frame of $\phi$. Note that it is the correlation
between $P_{s}^{y}(\mathbf{p})$ and $P_{\bar{s}}^{y}(-\mathbf{p})$
\cite{Efremov:1981vs} that is essential to resolve the puzzle in
$\rho_{00}^{\phi}$. Inserting (\ref{eq:py-s-sbar}) into (\ref{eq:rho00})
and taking an average of $\rho_{00}^{\phi}(t,\mathbf{x})$ over the
fireball volume $V$ and the polarization time $t$ with an effective
temperature $T_{\mathrm{eff}}$, we obtain 
\begin{eqnarray}
\rho_{00}^{\phi} & \approx & \frac{1}{3}-\frac{4}{9}\left\langle P_{\bar{\Lambda}}^{y}P_{\Lambda}^{y}\right\rangle +\frac{1}{27m_{s}^{2}}\left\langle \mathbf{p}^{2}\right\rangle _{\phi}\left\langle \varepsilon_{z}^{2}+\varepsilon_{x}^{2}\right\rangle \nonumber \\
 &  & -\frac{e^{2}}{243m_{s}^{4}T_{\mathrm{eff}}^{2}}\left\langle \mathbf{p}^{2}\right\rangle _{\phi}\left\langle E_{z}^{2}+E_{x}^{2}\right\rangle ,\label{eq:rho-00-phi-3}
\end{eqnarray}
where we have used $\left\langle \mathbf{p}\right\rangle _{\phi}=0$,
$\left\langle p_{z,x}^{2}\right\rangle _{\phi}=(1/3)\left\langle \mathbf{p}^{2}\right\rangle _{\phi}$,
$\left\langle p_{z}p_{x}\right\rangle _{\phi}=0$, with $\left\langle a(\mathbf{p})\right\rangle _{\phi}\equiv(2\pi)^{-3}\int d^{3}\mathbf{p}\left|\psi_{\phi}(\mathbf{p})\right|^{2}a(\mathbf{p})$
being the mean value of a momentum function $a(\mathbf{p})$ in the
$\phi$ meson wave function in momentum space, and replaced $T$ by
the effective temperature $T_{\mathrm{eff}}$ of the fireball. From
the $\phi$ meson wave function in the quark potential model \cite{SilvestreBrac:1996bg,Hiyama:2003cu},
we have $\left\langle \mathbf{p}^{2}\right\rangle _{\phi}\approx0.18\:\mathrm{GeV}^{2}\approx9.45m_{\pi}^{2}$
with $m_{\pi}\approx(2/3)m_{\pi^{\pm}}+(1/3)m_{\pi^{0}}\approx$138.05
MeV. Using Eq. (\ref{eq:p-lambda}), the second term in the right-hand
side of Eq. (\ref{eq:rho-00-phi-3}) is denoted as $c_{\Lambda}\equiv-(4/9)\left\langle P_{\bar{\Lambda}}^{y}P_{\Lambda}^{y}\right\rangle $:
\begin{equation}
c_{\Lambda}=-\frac{1}{9}\left\langle \omega_{y}^{2}\right\rangle +\frac{Q_{s}^{2}}{9m_{s}^{2}T_{\mathrm{eff}}^{2}}\left\langle B_{y}^{2}\right\rangle .\label{eq:lambda-lambda-bar}
\end{equation}
We see that the contribution to $\rho_{00}^{\phi}$ from the vorticity
is always negative while that from the magnetic field is always positive.
We also see that the magnitudes of $\left\langle \omega_{y}^{2}\right\rangle $
and $\left\langle B_{y}^{2}\right\rangle $ are constrained by the
data of $P_{\Lambda}$ and $P_{\bar{\Lambda}}$, but this is not the
case for $\left\langle \varepsilon_{z}^{2}+\varepsilon_{x}^{2}\right\rangle $
and $\left\langle E_{z}^{2}+E_{x}^{2}\right\rangle $ in Eq. (\ref{eq:rho-00-phi-3}). 

We denote the third and fourth term in the right-hand side of Eq.
(\ref{eq:rho-00-phi-3}) as $c_{\varepsilon}$ and $c_{E}$ respectively.
Note that all these terms are either positive or negative definite,
which is a good feature of $\rho_{00}^{\phi}$. The $c_{\Lambda}$
term has two contributions: the vorticity contribution is negative
and the magnetic field contribution is positive, they are all in the
order $10^{-3}$ to $10^{-4}$ according to the simulations using
hydrodynamic models \cite{Pang:2012he,Pang:2018zzo} and transport
models \cite{Voronyuk:2011jd,Toneev:2011aa}, respectively. The $c_{\varepsilon}$
term provides a positive contribution to $\rho_{00}^{\phi}$ but not
constrained by the data of $\Lambda$ polarization. This term comes
from the fluid vorticity and can be estimated by the hydrodynamic
simulation. We use CLVisc \cite{Pang:2012he,Pang:2018zzo}, a (3+1)D
viscous hydrodynamic model, to calculate $\left\langle \varepsilon_{z}^{2}+\varepsilon_{x}^{2}\right\rangle $
at the freezeout. The numerical results show $\left\langle \varepsilon_{z}^{2}+\varepsilon_{x}^{2}\right\rangle \sim10^{-4}$.
Using the constituent quark mass for $m_{s}$ of about 450 MeV, $c_{\varepsilon}$
is even more suppressed. The $c_{E}$ term is from the electric field
which is also absent in the $\Lambda$ polarization (\ref{eq:p-lambda})
and therefore not constrained by the data of $\Lambda$ polarization.
The peak value for $eE\equiv e\sqrt{\left\langle E_{z}^{2}+E_{x}^{2}\right\rangle }$
is about $m_{\pi}^{2}$ according to the simulation based on the Parton-Hadron-String
Dynamics (PHSD) transport model \cite{Cassing:2009vt} which includes
a dynamical generation of retarded electromagnetic fields \cite{Voronyuk:2011jd,Toneev:2011aa},
where we set $m_{s}\approx450$ MeV and $T_{\mathrm{eff}}\approx$100-300
MeV for Au+Au collisions in the collision energy range 20-200 GeV.
Then we obtain $c_{E}\sim10^{-5}$, which cannot give a large deviation
of $\rho_{00}^{\phi}$ from $1/3$. 

\section{Spin polarization in a meson field of $\phi$}

Like the electromagnetic field, a mean field of the $\phi$ meson,
if exists, can also polarize $s$ and $\bar{s}$ and contribute to
$\rho_{00}^{\phi}$. The role of the mean field of vector mesons in
the polarization of the Lambda hyperon was proposed in Ref. \cite{Csernai:2018yok}.
The electric and magnetic part of the $\phi$ meson field $\mathbf{E}_{\phi}$
and $\mathbf{B}_{\phi}$ can be obtained by the field potential $\phi^{\mu}$
in the same way as for the electromagnetic field: $F_{\phi}^{\mu\nu}=\partial^{\mu}\phi^{\nu}-\partial^{\nu}\phi^{\mu}$.
This is in analogy with the vector dominance model \cite{Sakurai:1960ju,Bauer:1977iq}.
Similar to the meson field out of the baryon current in Ref. \cite{Csernai:2018yok},
$\phi^{\mu}$ can be approximately proportional to the current density
of the strangeness quantum number, $\phi^{\mu}\approx-(g_{\phi}/m_{\phi}^{2})J_{s}^{\mu}$,
known as the current-field identity \cite{GellMann:1961tg,Kroll:1967it}
in the vector dominance model \cite{Sakurai:1960ju,Bauer:1977iq}.
Here $m_{\phi}$ is the $\phi$ meson mass, and $g_{\phi}$ is the
coupling constant of the $s$ quark to the $\phi$ meson in the quark-meson
model \cite{Zacchi:2015lwa,Zacchi:2016tjw}. 

Note that the contribution from $s$ and $\bar{s}$ to $J_{s}^{\mu}$
is negative and positive, respectively. The strangeness current density
in the central rapidity region is assumed to be a function of time
and space 
\begin{equation}
J_{s}^{\mu}(t,\mathbf{x})=(\rho_{s},\mathbf{J}_{s})=(\rho_{s},j_{s}^{x},j_{s}^{y},j_{s}^{z}),\label{eq:current-density-1}
\end{equation}
It must satisfy strangeness conservation $\partial_{\mu}J_{s}^{\mu}=0$
with the condition $\int d^{3}\mathbf{x}\rho_{s}(t,\mathbf{x})=0$.
The electric and magnetic part of the $\phi$ field that contribute
to the spin alignment along $+y$ direction are given by
\begin{eqnarray}
\mathbf{E}_{\phi} & = & \hat{\mathbf{z}}\frac{g_{\phi}}{m_{\phi}^{2}}\tilde{E}_{\phi}^{z}+\hat{\mathbf{x}}\frac{g_{\phi}}{m_{\phi}^{2}}\tilde{E}_{\phi}^{x},\nonumber \\
\mathbf{B}_{\phi} & = & \hat{\mathbf{y}}\frac{g_{\phi}}{m_{\phi}^{2}}\left(\frac{\partial j_{s}^{z}}{\partial x}-\frac{\partial j_{s}^{x}}{\partial z}\right),
\end{eqnarray}
where $\tilde{E}_{\phi}^{i}=\tilde{E}_{\phi,i}\equiv\nabla_{i}\rho_{s}+\partial j_{s}^{i}/\partial t$
with $i=x,y,z$. The $z$ component of $\mathbf{J}_{s}$ in (\ref{eq:current-density-1})
is the result of the difference in the parton distribution function
for $s$ and $\bar{s}$ in nucleons: $s(x_{B})\neq\bar{s}(x_{B})$
in different regions of $x_{B}$, where $x_{B}$ is the momentum fraction
(Bjorken variable) carried by $s$ and $\bar{s}$ in the proton. Although
the uncertainty in extracting $s(x_{B})$ and $\bar{s}(x_{B})$ in
the nucleon sea from experimental data \cite{Bazarko:1994tt,Rabinowitz:1993xx,Arneodo:1996qe}
is large, there are strong evidences \cite{Arneodo:1996qe,Boros:1998qh}
for $s(x_{B})\neq\bar{s}(x_{B})$. Extensive theoretical studies have
been done on the asymmetry of $s(x_{B})$ and $\bar{s}(x_{B})$ in
the past 30 years \cite{Signal:1987gz,Brodsky:1996hc,Holtmann:1996be,Christiansen:1998dz,Cao:1999da,Cao:2003ny,Ding:2004ht,Traini:2013zqa,Vega:2015hti}.
In nucleus-nucleus collisions, this leads to a non-zero strangeness
current $j_{s}^{z}$ which may depend on time. We have also generalized
this feature by introducing $\rho_{s}$, $j_{s}^{x}$ and $j_{s}^{y}$
in Eq. (\ref{eq:current-density-1}). 

Then the contribution from the $\phi$ meson field can be obtained
from Eqs. (\ref{eq:py-s-sbar},\ref{eq:p-lambda},\ref{eq:rho-00-phi-3},\ref{eq:lambda-lambda-bar})
by replacements: $\mathbf{B}\rightarrow\mathbf{B}_{\phi}$, $\mathbf{E}\rightarrow\mathbf{E}_{\phi}$
and $Q_{s}=-\frac{1}{3}e\rightarrow g_{\phi}$. Now $P_{s/\bar{s}}^{y}$
in Eq. (\ref{eq:py-s-sbar}) have two additional terms: $\pm g_{\phi}B_{\phi}^{y}/(2m_{s}T)$
and $\pm g_{\phi}\hat{\mathbf{y}}\cdot\left(\mathbf{E}_{\phi}\times\mathbf{p}_{s/\bar{s}}\right)/(2m_{s}^{2}T)$.
Correspondingly, $P_{\Lambda/\bar{\Lambda}}^{y}(t,\mathbf{x})$ in
(\ref{eq:p-lambda}) has an additional term $\pm g_{\phi}B_{\phi}^{y}/(2m_{s}T)$
which is constrained by the data. We see that it is $\mathbf{B}_{\phi}$
instead of $\mathbf{E}_{\phi}$ that contributes to $P_{\Lambda/\bar{\Lambda}}^{y}(t,\mathbf{x})$.
Equation (\ref{eq:rho-00-phi-3}) becomes 
\begin{equation}
\rho_{00}^{\phi}\approx\frac{1}{3}+c_{\Lambda}+c_{\varepsilon}+c_{E}+c_{\phi},\label{eq:rho-00-phi-2}
\end{equation}
where $c_{\phi}$ is from the $\phi$ field 
\begin{equation}
c_{\phi}\equiv\frac{g_{\phi}^{2}}{27m_{s}^{2}T_{\mathrm{eff}}^{2}}\left[3\left\langle B_{\phi,y}^{2}\right\rangle -\frac{\left\langle \mathbf{p}^{2}\right\rangle _{\phi}}{m_{s}^{2}}\left\langle E_{\phi,z}^{2}+E_{\phi,x}^{2}\right\rangle \right].\label{eq:c-phi}
\end{equation}
Note that the average is taken over the space-time volume. In deriving
(\ref{eq:rho-00-phi-2}) we have assumed that there are no correlations
among different fields (fluid field, electromagnetic field, $\phi$
field), e.g. between fluid and electromagnetic field, between $\mathbf{B}$
and $\mathbf{B}_{\phi}$, and between $\mathbf{E}$ and $\mathbf{E}_{\phi}$,
etc.. We have also assumed that there is no correlation between the
electric and magnetic part of the same field. The important feature
of Eq. (13) is that $c_{\phi}$ has positive contribution from $\mathbf{B}_{\phi}$
and negative contribution from $\mathbf{E}_{\phi}$ in the form of
field squares which are not constrained by $P_{\Lambda/\bar{\Lambda}}^{y}(t,\mathbf{x})$.
We note that Eq. (\ref{eq:rho-00-phi-2}) is for $\rho_{00}^{\phi}$
in the $y$ direction, one can obtain $\rho_{00}^{\phi}$ in the $x$
or $z$ direction as well. For $\rho_{00}^{\phi}$ in the $x$ direction,
one can just replace $\omega_{y}$, $B_{y}$ and $B_{\phi}^{y}$ in
$c_{\Lambda}$ by $\omega_{x}$, $B_{x}$ and $B_{\phi}^{x}$ respectively,
and replace $\varepsilon_{x}$, $E_{x}$ and $E_{\phi}^{x}$ in $c_{\varepsilon}$,
$c_{E}$ and $c_{\phi}$ by $\varepsilon_{y}$, $E_{y}$ and $E_{\phi}^{y}$
respectively. 

As we have shown in Sec. \ref{sec:em} that $c_{\Lambda}$, $c_{\varepsilon}$
and $c_{E}$ in Eq. (\ref{eq:rho-00-phi-2}) are negligibly small
compared with $1/3$ for Au+Au collisions in the collision energy
range 20-200 GeV. If the data show that $\rho_{00}^{\phi}$ is larger
than $1/3$ by at least a few percent, according to our model, the
deviation may possibly be from $c_{\phi}$ involving the magentic
part of the $\phi$ field. A good feature of $\rho_{00}^{\phi}$ is
that each contribution is in square up to a sign, so it is either
positive or negative definite. This property does not depend on the
procedure of taking an average or on choices of parameters. It exists
even for fluctuating fields (the vorticity, electromagnetic and $\phi$
field). Therefore $\rho_{00}^{\phi}$ is a good analyzer for fields
even if they may fluctuate strongly in space-time. 

\begin{figure}
\caption{The spin matrix element $\rho_{00}$ for the $\phi$ meson in heavy-ion
collisions from Eq. (\ref{eq:rho-00-phi-2}). The thin horizontal
solid line shows the no-alignment value $\rho_{00}=1/3$. Three values
of $G_{s}^{(y)}$ are chosen. \label{fig:rho-00-figure}}

\includegraphics[scale=0.6]{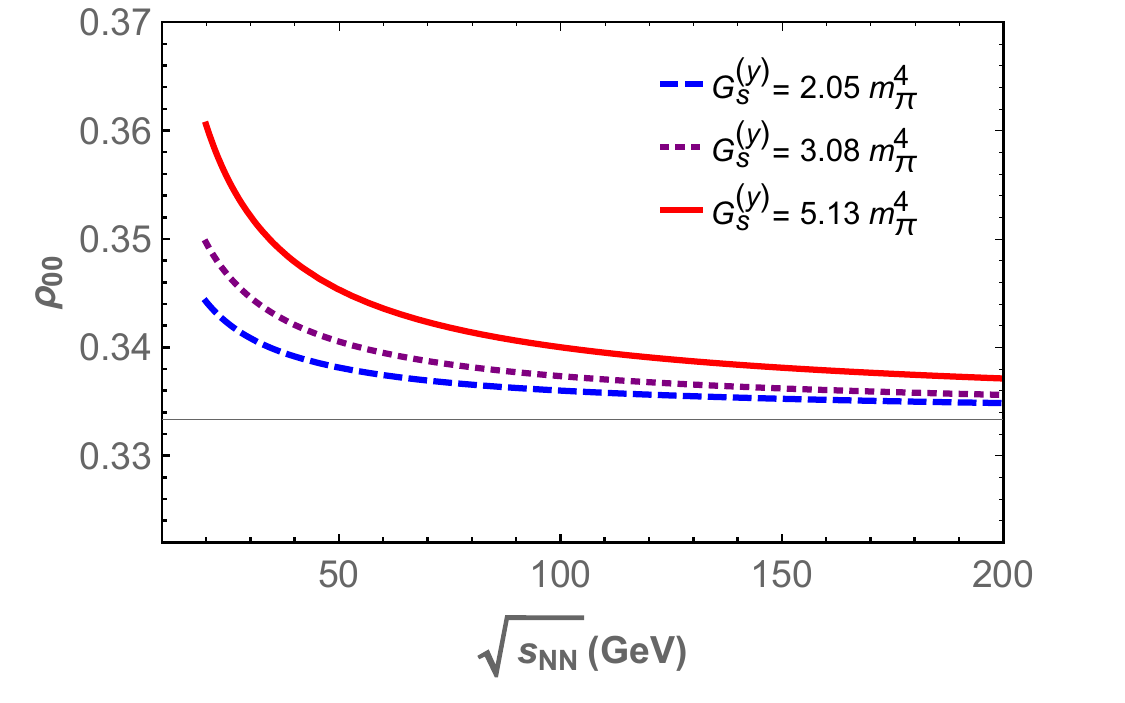}
\end{figure}

We can estimate in a simple model the dominant contribution to $\rho_{00}^{\phi}$
from the last term of Eq. (\ref{eq:rho-00-phi-2}). We choose the
effective temperature as $T_{\mathrm{eff}}\propto\tau_{0}^{-1/3}\left(dn_{\mathrm{ch}}/d\eta\right)_{\eta=0}^{1/3},$
where $\tau_{0}\sim s_{NN}^{-1/2}$ and $(dn_{\mathrm{ch}}/d\eta)_{\eta=0}\propto-0.4+0.39\ln s_{NN}$
is the pseudorapidity density of charged particles at the central
pseudorapidity $\eta=0$ and the collision energy $s_{NN}^{1/2}$
should take the dimensionless number when expressed in the unit GeV
\cite{Back:2005hs}. We set $T_{\mathrm{eff}}=300$ MeV at $s_{NN}^{1/2}=200$
GeV for calibration. In this way the collision energy behavior of
$\rho_{00}^{\phi}$ is solely from $T_{\mathrm{eff}}$ which is a
strong assumption in this order of magnitude estimate. As an approximation,
we assume that $\partial\mathbf{J}_{s}/\partial t$ and $\nabla\times\mathbf{J}_{s}$
do not depend on the collision energy. We set the values of the following
parameters: $m_{s}=$450 MeV and $G_{s}^{(y)}=(2.05,3.08,5.13)\:m_{\pi}^{4}$
where $G_{s}^{(y)}\equiv g_{\phi}^{2}\left[3\left\langle B_{\phi,y}^{2}\right\rangle -\left(\left\langle \mathbf{p}^{2}\right\rangle _{\phi}/m_{s}^{2}\right)\left\langle E_{\phi,z}^{2}+E_{\phi,x}^{2}\right\rangle \right]$.
Note that the value of $g_{\phi}$ can be taken from the constraint
by the compact star properties in the quark-meson model \cite{Zacchi:2015lwa,Zacchi:2016tjw}.
With these values of parameters the dominant contribution to $\rho_{00}^{\phi}$,
$c_{\phi}$ in Eq. (\ref{eq:rho-00-phi-2}), as a function of collision
energy in Au+Au collisions is shown in Fig. \ref{fig:rho-00-figure}.
We see in Fig. \ref{fig:rho-00-figure} that $\rho_{00}^{\phi}$ decreases
with the collision energy.

A natural question arises: are the theory and conclusion in this paper
valid for another vector meson $K^{*0}(892)$? The answer would be
no. There are a few reasons for it. First, due to unequal masses of
$\bar{s}$ and $d$, one cannot derive similar formula to Eq. (\ref{eq:rho-00-phi-3})
in which terms of vorticity and those of electric and magnetic field
are decoupled. Therefore one cannot build up a simple relationship
between $\rho_{00}$ for $K^{*0}$, $\rho_{00}^{K^{*0}}$, and the
hyperon polarization. In $\rho_{00}^{K^{*0}}$ each contribution can
be either positive or negative, so it is not easy to single out a
specific contribution from $\rho_{00}$ which belongs to the vorticity,
electromagnetic field or mesonic filed without ambiguity. Second,
the interaction of $K^{*0}$ with the surrounding matter is much stronger
than the $\phi$ meson. In this sense, the $\phi$ meson is a cleaner
probe than $K^{*0}$ to the state of the fireball. Actually, preliminary
data from the ALICE experiment show that $\rho_{00}$ for the $K^{*0}$
meson is less than $1/3$ at LHC energies \cite{BedangadasMohantyfortheALICE:2017xgh,Singh:2018uad},
which is very different from the $\phi$ meson. Another question is:
what happens for $\rho_{00}$ at LHC energies? From the energy behavior
in Fig. \ref{fig:rho-00-figure}, we expect that negative $c_{\Lambda}$
and $c_{\varepsilon}$ would be comparable to positive $c_{\phi}$
at LHC energies. In this case, whether $\rho_{00}$ is larger or smaller
than 1/3 depends on a fine-tuning of each terms.

\section{Summary}

Due to the difference in the parton distribution function of $s$
and $\bar{s}$ in high energy proton-proton collisions, the longitudinal
momenta carried by $s$ and $\bar{s}$ are not equal. This leads to
a non-vanishing collective strangeness current in the beam direction
in high energy heavy-ion collisions. We generalize this feature to
transverse directions. Such a strangeness current gives rise to a
non-vanishing electric and magnetic part of the mean $\phi$ field,
$\mathbf{E}_{\phi}$ and $\mathbf{B}_{\phi}$, respectively. Like
the magnetic field, $\mathbf{B}_{\phi}$ can also polarize $s$ and
$\bar{s}$ through their magnetic moments which contributes to the
polarization of $\Lambda$ and $\bar{\Lambda}$, while the contribution
from $\mathbf{E}_{\phi}$ is absent and therefore is not constrained
by the polarization of $\Lambda$ and $\bar{\Lambda}$. Note that
$\mathbf{E}_{\phi}$ can also polarize $s$ and $\bar{s}$ through
the spin-orbit coupling, the same coupling that is responsible for
the nuclear shell structure at the nucleon level. The contributions
from $\mathbf{B}_{\phi}^{2}$ and $\mathbf{E}_{\phi}^{2}$ to $\rho_{00}^{\phi}$
are positive and negative respectively, if the polarization of $s$
and $\bar{s}$ is assumed to be only along the OAM direction ($y$
direction) and if $\phi$ mesons are static (with vanishing momenta).
Both $\mathbf{B}_{\phi}^{2}$ and $\mathbf{E}_{\phi}^{2}$ are not
constrained by the polarization data of $\Lambda$ and $\bar{\Lambda}$.
We then propose that a significant deviation of $\rho_{00}^{\phi}$
from 1/3 could indicate the presence of the $\phi$ field in heavy-ion
collisions which polarizes $s$ and $\bar{s}$ in the same way as
the electromagnetic field. The contributions are significant even
for fluctuating fields. In this sense $\rho_{00}^{\phi}$ is a good
analyzer for fluctuating fields.

\section*{Acknowledgments}

The authors thank E. Bratkovskaya, L.-G. Pang, D. Rischke, A.-H. Tang,
H.-Z. Wu for insightful discussions. QW thanks Q. Meng for providing
the $\phi$ meson wave function in the quark potential model. QW is
supported in part by the National Natural Science Foundation of China
(NSFC) under Grant No. 11535012 and 11890713. XLS and LO are supported
by the Deutsche Forschungsgemeinschaft (DFG, German Research Foundation)
through the Collaborative Research Center CRC-TR 211\textquotedblleft Strong-interaction
matter under extreme conditions\textquotedblright{} with Project No.
315477589. LO is supported by the Alexander von Humboldt-Stiftung.

\appendix

\section{Note added after publication}

The main result in Eq. (\ref{eq:c-phi}) is derived originally from
Eq. (\ref{eq:rho00}) based on two assumptions or conditions: (a)
$\phi$ mesons are static with vanishing momenta; (b) The polarization
of $s$ and $\overline{s}$ quarks is only along the OAM direction,
i.e. the $y$ direction. We can relax the condition (b) by allowing
the polarization of $s$ and $\overline{s}$ quarks can be along all
directions but the spin quantization direction is still in the $y$
direction. In this case Eq. (\ref{eq:rho00}) becomes 
\begin{eqnarray}
\rho_{00}^{\phi}(t,\mathbf{x}) & \approx & \frac{1}{3}-\frac{4}{9}\int\frac{d^{3}\mathbf{p}}{(2\pi)^{3}}\left|\psi_{\phi}(\mathbf{p})\right|^{2}\nonumber \\
 &  & \times\left\{ P_{s}^{y}(\mathbf{p})P_{\bar{s}}^{y}(-\mathbf{p})-\frac{1}{2}\left[P_{s}^{x}(\mathbf{p})P_{\bar{s}}^{x}(-\mathbf{p})+P_{s}^{z}(\mathbf{p})P_{\bar{s}}^{z}(-\mathbf{p})\right]\right\} .\label{eq:rho-00-3-direction}
\end{eqnarray}
The polarization of $s$ and $\overline{s}$ quarks can be written
as 
\begin{eqnarray}
P_{s/\bar{s}}(t,\mathbf{x},\mathbf{p}_{s/\bar{s}}) & \approx & \frac{1}{2}\boldsymbol{\omega}+\frac{1}{2m_{s}}(\boldsymbol{\varepsilon}\times\mathbf{p}_{s/\bar{s}})\pm\frac{Q_{s}}{2m_{s}T}\mathbf{B}\pm\frac{Q_{s}}{2m_{s}^{2}T}\left(\mathbf{E}\times\mathbf{p}_{s/\bar{s}}\right)\nonumber \\
 &  & \pm\frac{g_{\phi}}{2m_{s}T}\mathbf{B}_{\phi}\pm\frac{g_{\phi}}{2m_{s}^{2}T}\left(\mathbf{E}_{\phi}\times\mathbf{p}_{s/\bar{s}}\right),\label{eq:py-s-sbar-1}
\end{eqnarray}
which has three contributions: the vorticity, electromagnetic, and
$\phi$ fields. Inserting (\ref{eq:py-s-sbar-1}) into (\ref{eq:rho-00-3-direction})
we obtain $\rho_{00}^{\phi}$ for static $\phi$ mesons
\begin{eqnarray}
\rho_{00}^{\phi} & \approx & \frac{1}{3}-\frac{1}{9}\left[\left\langle \omega_{y}^{2}\right\rangle -\frac{1}{2}\left\langle \omega_{x}^{2}+\omega_{z}^{2}\right\rangle \right]\nonumber \\
 &  & -\frac{\left\langle {\bf {\bf p}}^{2}\right\rangle _{\phi}}{27m_{s}^{2}}\left[\left\langle \varepsilon_{y}^{2}\right\rangle -\frac{1}{2}\left\langle \varepsilon_{x}^{2}+\varepsilon_{z}^{2}\right\rangle \right]\nonumber \\
 &  & +\frac{Q_{s}^{2}}{9m_{s}^{2}T_{\mathrm{eff}}^{2}}\left[\left\langle B_{y}^{2}\right\rangle -\frac{1}{2}\left\langle B_{x}^{2}+B_{z}^{2}\right\rangle \right]\nonumber \\
 &  & +\frac{Q_{s}^{2}\left\langle {\bf {\bf p}}^{2}\right\rangle _{\phi}}{27m_{s}^{4}T_{\mathrm{eff}}^{2}}\left[\left\langle E_{y}^{2}\right\rangle -\frac{1}{2}\left\langle E_{x}^{2}+E_{z}^{2}\right\rangle \right]\nonumber \\
 &  & +\frac{g_{\phi}^{2}}{9m_{s}^{2}T_{\mathrm{eff}}^{2}}\left[\left\langle B_{\phi,y}^{2}\right\rangle -\frac{1}{2}\left\langle B_{\phi,x}^{2}+B_{\phi,z}^{2}\right\rangle \right]\nonumber \\
 &  & +\frac{g_{\phi}^{2}\left\langle {\bf {\bf p}}^{2}\right\rangle _{\phi}}{27m_{s}^{4}T_{\mathrm{eff}}^{2}}\left[\left\langle E_{\phi,y}^{2}\right\rangle -\frac{1}{2}\left\langle E_{\phi,x}^{2}+E_{\phi,z}^{2}\right\rangle \right].
\end{eqnarray}
If $\rho_{00}^{\phi}$ is dominated by the $\phi$ field, we obtain
$\rho_{00}^{\phi}$ for the spin quantization direction being in the
$y$ direction as 
\begin{equation}
\rho_{00}^{\phi}\approx\frac{1}{27m_{s}^{2}T_{\mathrm{eff}}^{2}}G_{s}^{(y)},
\end{equation}
where $G_{s}^{(y)}$ now becomes 
\begin{eqnarray}
G_{s}^{(y)} & = & \frac{3}{2}Q_{yy}(\mathbf{B}_{\phi})-\frac{\left\langle {\bf {\bf p}}^{2}\right\rangle _{\phi}}{2m_{s}^{2}}Q_{yy}(\mathbf{E}_{\phi})\nonumber \\
 & = & g_{\phi}^{2}\left[3\left\langle B_{\phi,y}^{2}\right\rangle +\frac{\left\langle {\bf {\bf p}}^{2}\right\rangle _{\phi}}{m_{s}^{2}}\left\langle E_{\phi,y}^{2}\right\rangle \right.\nonumber \\
 &  & \left.-\frac{3}{2}\left\langle B_{\phi,x}^{2}+B_{\phi,z}^{2}\right\rangle -\frac{\left\langle {\bf {\bf p}}^{2}\right\rangle _{\phi}}{2m_{s}^{2}}\left\langle E_{\phi,x}^{2}+E_{\phi,z}^{2}\right\rangle \right].\label{eq:cy}
\end{eqnarray}
Here we have defined the average quadrupole moments for $\mathbf{B}_{\phi}$
and $\mathbf{E}_{\phi}$ as 
\begin{eqnarray}
Q_{ij}(\mathbf{B}_{\phi}) & \equiv & g_{\phi}^{2}\left\langle 3\mathbf{B}_{\phi,i}\mathbf{B}_{\phi,j}-\left|\mathbf{B}_{\phi}\right|^{2}\right\rangle ,\nonumber \\
Q_{ij}(\mathbf{E}_{\phi}) & \equiv & g_{\phi}^{2}\left\langle 3\mathbf{E}_{\phi,i}\mathbf{E}_{\phi,j}-\left|\mathbf{E}_{\phi}\right|^{2}\right\rangle ,
\end{eqnarray}
with $i,j=x,y,z$. We see in Eq. (\ref{eq:cy}) that if the spin quantization
direction is chosen to be the $y$ direction, the positive contributions
to $\rho_{00}^{\phi}$ come from $B_{\phi,y}$ and $E_{\phi,y}$ while
negative ones to $\rho_{00}^{\phi}$ come from $B_{\phi,i}$ and $E_{\phi,i}$
with $i=x,z$. We can take the difference between $\rho_{00}^{\phi}$
for the spin quantization in the $y$ (out-plane) direction from that
in the $x$ (in-plane) direction as 
\begin{eqnarray}
\Delta\rho_{00}^{\phi} & \equiv & \frac{1}{27m_{s}^{2}T_{\mathrm{eff}}^{2}}\left[G_{s}^{(y)}-G_{s}^{(x)}\right]\nonumber \\
 & \approx & \frac{g_{\phi}^{2}}{18m_{s}^{2}T_{\mathrm{eff}}^{2}}\left[3\left\langle B_{\phi,y}^{2}-B_{\phi,x}^{2}\right\rangle +\frac{\left\langle {\bf {\bf p}}^{2}\right\rangle _{\phi}}{m_{s}^{2}}\left\langle E_{\phi,y}^{2}-E_{\phi,x}^{2}\right\rangle \right].
\end{eqnarray}
This can also be tested in future experiments.

\bibliographystyle{apsrev}
\bibliography{ref}

\end{document}